\documentclass[]{article}

\usepackage{color}
\usepackage[usenames,dvipsnames,svgnames,table]{xcolor}
\usepackage{graphicx}

\small

\begin{document}

\twocolumn

\begin{center}
\fboxrule0.02cm
\fboxsep0.4cm
\fcolorbox{blue}{AliceBlue}{\rule[-0.9cm]{0.0cm}{1.8cm}{\parbox{7.8cm}
{ \begin{center}
{\Large\em Perspective}

\vspace{0.2cm}

{\large\bf Dust trapping in \\ protoplanetary disks} 



\vspace{0.2cm}

{\large\em Nienke van der Marel}
\vspace{0.2cm}

\centering
\includegraphics[width=0.24\textwidth]{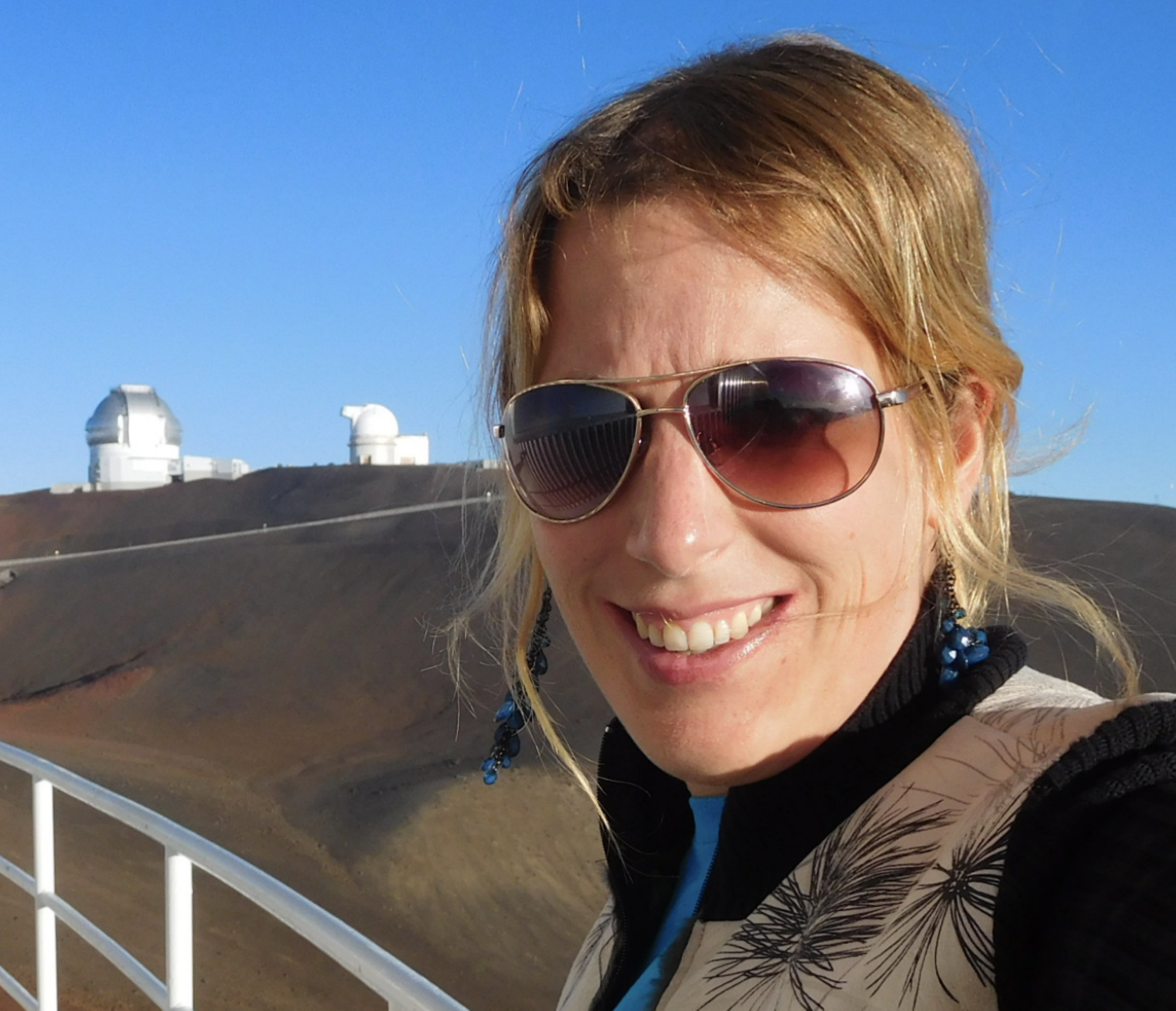}
\end{center}
}}}
\end{center}

\vspace{0.5cm}
\subsection{Introduction}
The planet formation process remains one of the major puzzles in modern-day astronomy. A large amount of research has been conducted in the second half of the twentieth century, but many questions remain unanswered. In particular, planet formation is hindered by a number of growth barriers, according to dust evolution theory, while observational evidence indicates that somehow these barriers must have been overcome. 

Planets are thought to form in protoplanetary disks of gas and dust around young stars, where the disk itself is a by-product of the star formation process due to conservation of angular momentum. Giant planets of $\sim$ Jupiter mass must form before the gas in the disk is dissipated, which is thought to happen within the first 5-10 million years of the life of the disk (Williams \& Cieza 2011). Although gravitational instability or fragmentation allow a rapid concentration and formation of Jupiter like planets (e.g. Helled et al. 2014 and references therein), protoplanetary disk masses are generally too low to make this mechanism a common pathway for the formation of planetary systems. This is a problem in particular considering the observed diversity in exoplanetary masses (e.g. Winn \& Fabrycky 2015), which cannot all be produced by fragmentation. The core accretion process, where the accretion of dust particles and planetesimals results into solid cores of $\sim$Earth mass, followed by runaway gas accretion, is a more promising mechanism to form the range of planets that are seen in both our Solar System and beyond. However, the core accretion process requires the growth of planetesimals from submicron-sized dust grains that are seen in the interstellar medium (ISM). These first steps in dust growth in protoplanetary disks, where the dust evolution is governed by coagulation, fragmentation and the drag forces between dust and gas, have proven to be one of the most challenging parts in the planet formation process (e.g. Testi et al. 2014 and references therein). 

\subsection{Grain growth}
\begin{figure}[htb]
\begin{center}
\includegraphics[width=0.4\textwidth]{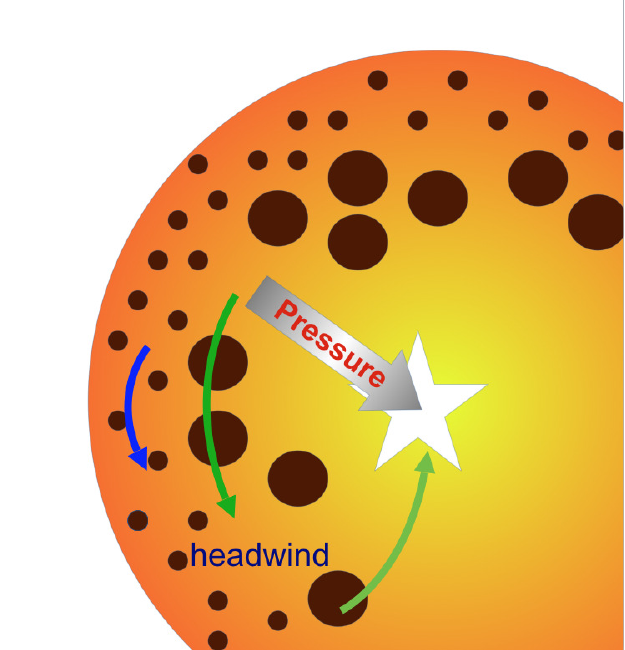}
\caption{Illustration of radial drift in a protoplanetary disk due to the pressure gradient in the gas. Small dust particles are tightly coupled with the gas (blue arrow), whereas larger dust particles are slowed down and spiral inward (green arrows) due to the surrounding gas at sub-Keplerian speed.}
\label{fig:drift}
\end{center}
\end{figure}

In the last two decades, clear evidence has been found for dust grain evolution in disks beyond the grain sizes in the interstellar medium, through analysis of mid-infrared spectra of silicate features (e.g. van Boekel et al., 2003; Kessler-Silacci et al., 2006) and (interferometric) millimeter observations of disks (e.g. Beckwith \& Sargent, 1991; Testi et al., 2003; Andrews \& Williams, 2005; Natta et al., 2007; Isella et al., 2009; Ricci et al., 2010), indicating the presence of millimeter and even centimeter-sized particles in the outer regions of protoplanetary disks. The main issue in explaining the presence of these dust particles is the so-called \emph{radial drift} problem, which results in a rapid inward drift of dust particles in a disk when grown to millimeter sizes (Whipple, 1972; Weidenschilling, 1977). For a disk with a smooth radial density profile, both the gas surface density and temperature, and thus the pressure, decrease radially outward (Figure \ref{fig:drift}). This additional pressure support results in a slightly sub-Keplerian orbital velocity for the gas. In contrast, dust particles are not pressure supported and want to orbit at Keplerian speed, but as they are embedded in a sub-Keplerian gas disk, they are forced to orbit at lower speed which leads to a removal of angular momentum from the particles to the gas, causing the particles to spiral inward. The dust particles are thus experiencing a drag force by the gas, depending on its Stokes number St, which describes the coupling of the particles to the gas and depends on the dust particle size and the local gas surface density (see e.g. Brauer et al., 2008; Birnstiel et al., 2010). For small dust particles, St$\ll$1, so they are strongly coupled to the gas and do not drift inwards, but radial drift becomes significant when the particle size increases and reaches its strongest value when St = 1. In practice, millimeter-dust particles in the outer disk can have St values close to unity and are expected to drift inwards on time scales as short as 100 years, hence further dust growth is hindered. An additional problem is \emph{fragmentation}: while low-velocity collisions lead to particle growth, high-velocity impacts lead to destruction according to experimental and theoretical work on dust interaction (Blum \& Wurm 2000). The combination of the radial drift and fragmentation problems is also called the `meter-size barrier' because a one-meter size object at 1 AU drifts efficiently inwards limiting further growth, and equivalently dust particles in the outer disk cannot grow beyond millimeter sizes.

\subsection{Pressure bumps}
A proposed solution to explain the presence of larger dust grains in disks as the start of planet formation, is a so-called \emph{dust trap}, where dust particles are being `trapped' in a long-lived pressure bump in the outer disk (Whipple, 1972; Klahr \& Henning, 1997; Rice et al., 2006; Brauer et al., 2008). Such a pressure bump can arise as a radial dust trap at the edges of dead zones (e.g. Varniere \& Tagger, 2006), at the edges of gas gaps cleared by planets (Zhu et al., 2012, Pinilla et al., 2012), in zonal flows (Johansen et al. 2009) or as azimuthal dust traps in long-lived vortices (e.g. Barge \& Sommeria, 1995; Klahr \& Henning, 1997), which can be the result of a Rossby Wave Instability of a radial pressure bump (e.g. Lovelace et al., 1999; Wolf \& Klahr, 2002; Lyra et al., 2009; Regaly et al., 2012). However, with limited spatial information on the distribution of millimeter dust grains in disks due to the low resolution images of disks, these proposed ideas remained mostly theoretical. Early continuum imaging by pioneering millimeter interferometers such as the SubMillimeter Array (SMA), Plateau de Bure Interferometer (PdBI) and the Combined Array for Research in Millimeter-wave Astronomy (CARMA) suggested the existence of lopsided disks (e.g. Brown et al., 2009; Isella et al., 2010), but the sensitivity, image fidelity and spatial resolution were insufficient for strong claims on the origin of the observed structures. The main targets of interest in these millimeter interferometry studies were the \emph{transition disks} --- disks where a deficit in the mid infrared part of their SED, indicating that the inner part of the disk was cleared of dust (see  Espaillat et al., 2014, for a review). The interferometric imaging was merely a confirmation that a dust cavity was indeed present and the disk was in fact ring-shaped. One of the explanations for these dust cavities is the presence of a recently formed young planet, that has cleared the material along its orbit, emptying the gap of dust and gas. However, at that time the link with dust trapping was only made due to the suggestion that some of the observed disk rings were somewhat asymmetric, which could not be explained by dynamical clearing alone. But due to the low signal-to-noise of the images, the measured asymmetries were barely significant and a link with dust traps remained highly debatable.

\begin{figure}[htb]
\begin{center}
\includegraphics[width=0.45\textwidth]{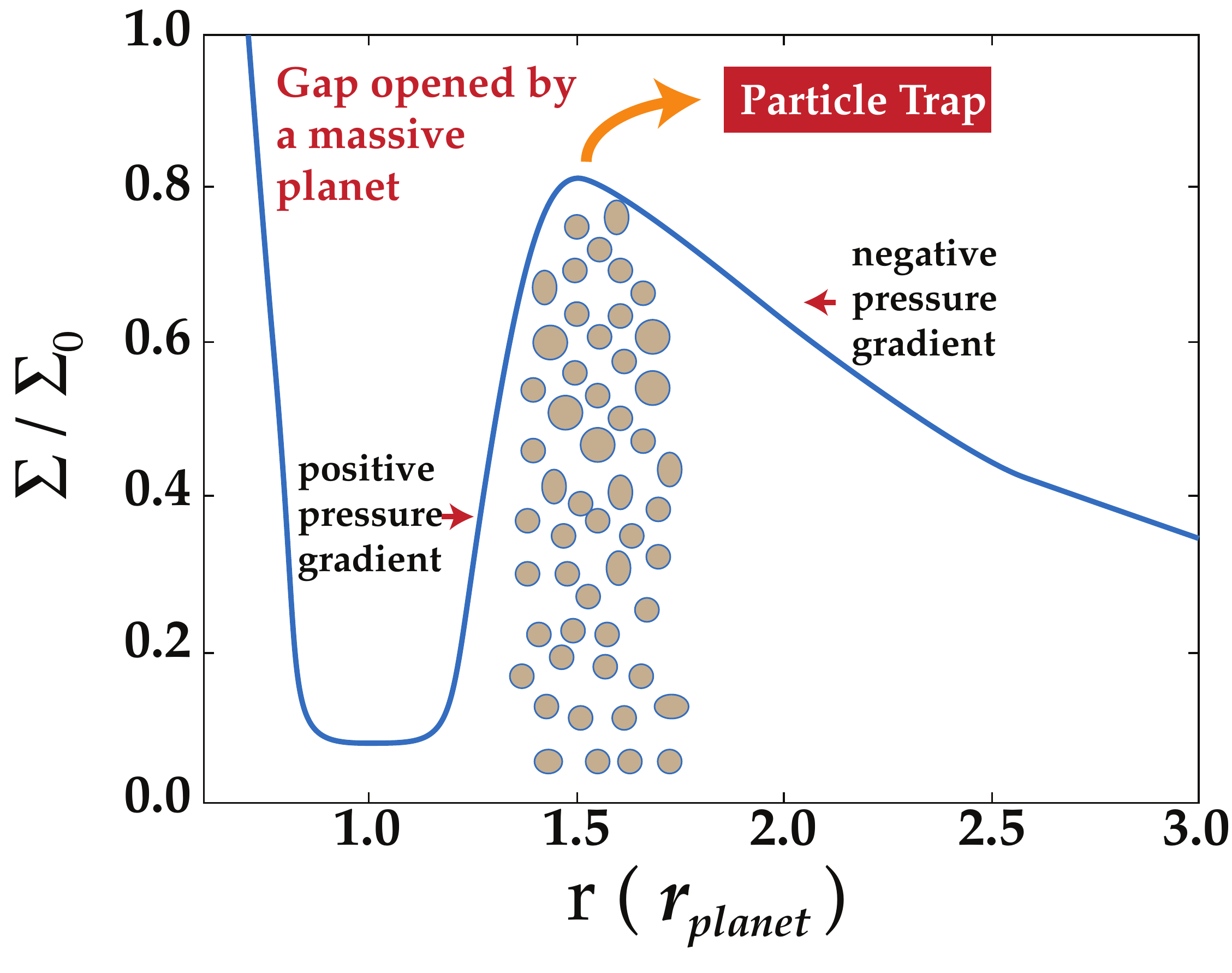}
\caption{Illustration of a pressure bump activing as a dust trap at the edge of a gap carved by a planet (figure by Paola Pinilla).}
\label{fig:dusttrap}
\end{center}
\end{figure}

\subsection{The discovery of the dust trap in Oph IRS 48}
\begin{figure}[htb]
\begin{center}
\includegraphics[width=0.5\textwidth]{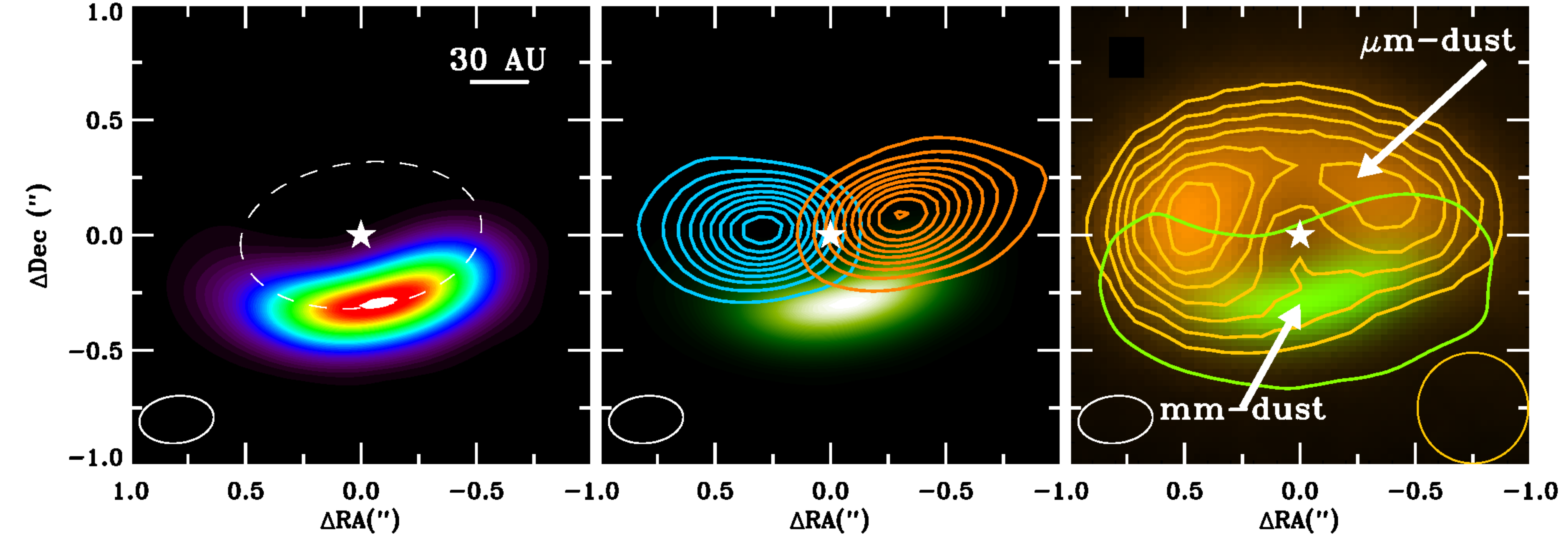}
\caption{Dust and gas observations of Oph IRS 48. From left to right: the ALMA 0.44mm continuum image; the integrated $^{12}$CO 6-5 line emission in red and blue and continuum in green; the ALMA continuum overlaid on the VLT/VISIR 18.7 $\mu$m image. Figure taken from van der Marel et al. 2013. }
\label{fig:irs48}
\end{center}
\end{figure}

The connection between theory and observations took a huge leap when the \emph{Atacama Large Millimeter/submillimeter Array} (ALMA) started its operations in 2011. Even with only one quarter of the current number of antennas, ALMA produced astonishing images of disks in Early Science Cycle 0. One of the Cycle 0 programs targeted Oph IRS 48, a bright transition disk around a Herbig star in the star forming region Ophiuchus, observed at the highest possible resolution at that time of $\sim$0.25" in the highest frequency Band 9 (690 GHz or 0.45 mm) (van der Marel et al., 2013). The aim of the program was to image the gas inside the cavity through the $^{12}$CO 6--5 line, motivated by mid infrared imaging of the thermal dust by VLT/VISIR at 18.7 $\mu$m (Geers et al., 2007), and spectro-astrometric results of VLT/CRIRES on the near infrared CO rovibrational line (Brown et al., 2012), both indicating that there was a cavity in gas \emph{and} dust, suggesting the presence of a companion. The ALMA continuum came for free in the spectral settings, but the team just expected to see a dust ring, similar to other transition disks. The actual continuum image was a big surprise, and at first very confusing; rather than a dust ring, the data showed an extreme asymmetry, with all continuum emission in a peanut-shape on the southern side of the star, whereas the gas, as traced by the CO emission, was evenly distributed and following a regular Keplerian butterfly pattern, with an inner gas cavity (Figure \ref{fig:irs48}). After consultation with dust evolution experts, it became clear that this was exactly what was predicted by theoretical models of a dust trap in a vortex: a solution to the radial drift problem! The crucial element was the combination of the highly asymmetric millimeter dust (an azimuthal contrast of $>$100), in combination with the ring-like CO emission, \emph{and} the axisymmetric distribution of the small dust grains as traced by the VISIR image . The images provided evidence for a scenario where a substellar/planetary companion has cleared its orbit in the gas and the small dust particles (seen in the $^{12}$CO emission and infrared images), the resulting pressure bump at the edge of the gap has become Rossby unstable, forming a long-lived vortex, trapping the millimeter dust particles in the azimuthal direction. As trapping is very efficient for larger dust particles, only a small (and undetectable) azimuthal contrast in the gas of the vortex is required to create the observed dust continuum contrast of $>$100  (de Val-Borro et al., 2007; Birnstiel et al., 2013). Oph IRS 48 thus became the first disk with clear evidence for the presence of dust trapping as an efficient way to explain millimeter grains and dust growth in protoplanetary disks. One major gap in this story though is the presence of the inner companion; in order to explain planet formation by dust traps, one needs another planet to begin with, creating a typical chicken-and-egg problem. However, as the companion is expected to be massive, potentially substellar, based on the depth of the gas gap (Bruderer et al. 2014), it may have formed as a binary companion rather than a planet. Regardless of the origin of the pressure bump, the evidence for the dust trap itself is solid, and thus a basis for dust trapping as a general phenomenon in protoplanetary disks.

\subsection{Other dust traps}
\begin{figure}[htb]
\begin{center}
\includegraphics[width=0.5\textwidth]{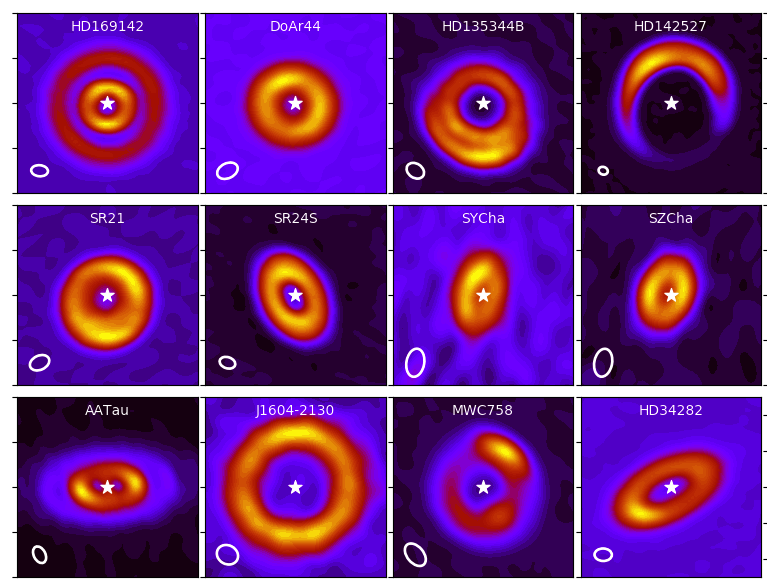}
\caption{Continuum overview of a number of transition disks observed by ALMA. The image reveals the large diversity of structures. Figure available for download at http://www.nienkevandermarel.com.}
\label{fig:overview}
\end{center}
\end{figure}

Several other transition disks were imaged in ALMA Early Science, showing a wide range of azimuthal asymmetry (and lack thereof), while SR~21 and HD~135344B (a.k.a. SAO~206462) were found to be somewhat lop-sided (Perez et al., 2014), and HD~142527 almost as asymmetric as Oph~IRS 48 (Casassus et al., 2013), other transition disks such as J1604-2130, LkCa15, DoAr44, and SR~24S were found to be perfectly circular (Zhang et al., 2014; van der Marel et al., 2015b; Pinilla et al., 2017). An overview of observed transition disk structures is shown in Figure \ref{fig:overview}. The wide range of asymmetries (and lack of extreme asymmetries) is no reason to dismiss the trapping mechanism; it is possible that in many cases only radial trapping is efficient. A vortex through Rossby wave instability is only expected to be long lived if the viscosity in the disk is low and the pressure bump is sufficiently steep, which could happen only in special cases or only on time scales of a fraction of the disk lifetime. Also eccentric gaps have been proposed to explain minor asymmetries in transition disk rings as an alternative to azimuthal trapping (Ataiee et al., 2014; Ragusa et al., 2017), although radial trapping in that case would still be happening. Radial trapping is observationally supported in many of these disks by comparison with the gas structure; the gas cavities as traced by CO isotopologues are always found to be smaller than the dust cavities (e.g. van der Marel et al., 2015b; Perez et al., 2015; van der Marel et al., 2016b; Boehler et al., 2017; Dong et al., 2017; Fedele et al., 2017; van der Marel et al., 2018), an indicator that the dust is trapped in a narrow ring at the outer edge of the cavity. Dust traps are further supported by scattered light images of the small dust grains by e.g. VLT/NACO and VLT/SPHERE (e.g. Garufi et al., 2013; Pinilla et al., 2015), where the millimeter dust grains are clearly located further out than the micrometer-sized dust grains, which can be explained in a similar way, often described as `dust filtration' (e.g. Rice et al., 2006; Zhu et al., 2012).
.

Multi-wavelength continuum imaging has allowed a further characterization of the dust trapping process. As trapping efficiency increases with particle size as the particle's Stokes number gets closer to unity, measuring the distribution of different particle sizes is a way to quantify a dust trap. The spectral index $\alpha_{mm}$ (millimeter flux $F_{\nu}\sim\nu^{\alpha_{mm}}$) provides information on the particle size in protoplanetary disks (Testi et al., 2014 and references therein). For (sub)micrometer-sized dust, such as found in the ISM, $\alpha_{mm}$ is typically 3.5-4.0, but when dust grows to millimeter sizes, $\alpha_{mm}$ is expected to decrease (Draine 2006; Ricci et al., 2010). When the dust emission is optically thin and in the Rayleigh-Jeans regime, the observable $\alpha_{mm}$ can be related to the dust opacity index $\beta = \alpha-2$, with $\beta<1$ for millimeter grains and related to the opacity as $\tau_{\nu}\propto\nu^{\beta}$. Spatially resolved multi-wavelength continuum observations thus provide a way to locate the regions where dust grains are growing, under the assumption of optically thin emission. Combining ALMA with VLA centimeter observations has demonstrated azimuthal dust trapping in Oph IRS 48 (van der Marel et al.,  2015a), HD142527 (Casassus et al., 2015) and MWC758 (Marino et al., 2015) as the azimuthal width decreases with wavelength. On the other hand, the asymmetry in AB Aur was shown to be \emph{more} extended at longer wavelengths (Fuente et al., 2017), opposite to the dust trapping predictions. This was interpreted as a time scale effect: as the vortex decays, dust particles diffuse on different time scales out of azimuthal trapping, depending on their particle size. Another interesting consequence of trapping in a vortex is a size segregation due to the vortex's self-gravity: whereas smaller grains will be trapped in the center of the vortex, larger grains are expected to be trapped \emph{ahead} in the azimuthal direction (Baruteau \& Zhu, 2016), which was indeed observed in the outer asymmetry in HD~135344B in multi-wavelength ALMA observations (Cazzoletti et al., subm.). Radial trapping was demonstrated in e.g. SR~21 and SR~24S (Pinilla et al., 2015; 2017) through spatially resolved $\alpha$ obtained from ALMA data, but optical depth is an issue: even in ALMA Band 7 (850$\mu$m), the dust emission is often found to be optically thick, and a low $\alpha$ merely indicates a change in optical depth rather than grain size. It has become clear that the optical depth is a real problem, as spatial resolution has further improved: disk rings are much narrower than initially assumed, pushing the emission to the optically regime even at millimeter wavelengths. Even marginally resolved observations indicate that millimeter emission in disk images may be dominated by optically thick emission (Tripathi et al., 2017).

Another spectacular ALMA result in this context is the imaging of disks at ultra high angular resolution of $\sim$20 mas or a few AU at the distance of nearby star forming regions. The mind-blowing image of HL~Tau (ALMA Consortium et al., 2015) of the ALMA Long Baseline Campaign has revealed that even dust disks without inner cavity may exist of ring-like structures. The HL~Tau image was quickly followed by other multi-ring disks, such as TW~Hya (Andrews et al., 2016) and HD~163296 (Isella et al., 2016). Also several transition disks turned out to consist of multiple rings, e.g. HD~97048 (van der Plas et al., 2017), HD~169142 (Fedele et al., 2017) or even a combination of rings and asymmetries in HD~135344B (van der Marel et al., 2016a), V1247 Ori (Kraus et al., 2017) and MWC~758 (Dong et al., 2018). An ALMA Large Program is on the way to reveal even more rings and other substructures in a large number of the brightest primordial disks (Andrews et al., in prep.). Explanations for the dust rings range from planets carving gaps (Lin \& Papaloizou, 1979), snow lines (Zhang et al., 2015), sintering (Okuzumi et al., 2016) and secular gravitational instabilities (Takahashi et al., 2016). If the dust rings are indeed caused by planets, trapping is expected to occur. Multi-wavelength observations indeed reveal radial variations of $\alpha_{mm}$ along the gaps and rings (Carrasco-Gonzalez et al., 2016; Tsukagoshi et al., 2016), but the evidence is still marginal with the currently available data. Snow lines could explain some of the ring locations without the need for trapping, but this suggests some correlation between the ring locations and the stellar properties which is not observed (van der Marel et al. in prep., Huang et al. in prep.). For these narrow rings, optical depth is an important issue in the interpretation of the dust emission. 

On the other hand, evidence for radial drift is evident as well in observations. SMA observations already revealed evidence for a segregation between the dust and gas in the IM Lup disk (Panic et al., 2009), with the gas outer radius being twice as large as that of the dust. Multi-wavelength continuum observations show a decrease of particle grain size with radius for a number of primordial disks (Perez et al., 2015; Tazzari et al., 2016). Clearly, in lack of one or more dust traps in the outer disk, dust particles do drift inwards to the nearest pressure maximum. On the other hand, this implies that any extended dust disk must in fact consist of one or more pressure bumps, to keep the dust particles away from the center of the disk.

\subsection{Future perspectives}
The discovery of dust traps in protoplanetary disks has revolutionized our understanding of planet formation. Although the origin of the pressure bumps remains an important question, their presence is almost indisputable. If caused by planets, the chicken-and-egg problem needs to be solved, in order to explain the presence of the first planet, although there is the interesting possibility of triggered planet formation where as soon as the first planet is formed, the next ones follow in sequence. Spatially resolved observations of the gas are crucial in order to understand the origin of the pressure bumps. A further characterization of the dust traps requires multi-wavelength continuum observations at very high spatial resolution, preferably at optically thin wavelengths. The Next Generation Very Large Array could play a crucial role here, when it can observe disks at centimeter wavelengths at the same level of detail as ALMA.

\footnotesize

\bibliographystyle{apj}
\bibliography{../../myrefs}

{\bf References:}\\
ALMA Partnership, A., et al. 2015, ApJ, 808, L3\\
Andrews, S. M., \& Williams, J. P. 2005, ApJ, 631, 1134\\
Andrews, S. M., et al. 2016, ApJ, 820, L40\\
Ataiee, S., et al. 2014, A\&A, 572, A61\\
Barge, P., \& Sommeria, J. 1995, A\&A, 295, L1\\
Baruteau, C., \& Zhu, Z. 2016, MNRAS, 458, 3927\\
Beckwith, S. V. W., \& Sargent, A. I. 1991, ApJ, 381, 250\\
Birnstiel, T., Dullemond, C. P., \& Brauer, F. 2010, A\&A, 513, A79 \\
Birnstiel, T., Dullemond, C. P., \& Pinilla, P. 2013, A\&A, 550, L8 \\
Blum, J., \& Wurm, G. 2000, Icarus, 143, 138\\
Boehler, Y., et al. 2017, ApJ, 840, 60\\
Brauer, F., Dullemond, C. P., \& Henning, T. 2008, A\&A, 480, 859 \\
Brown, J. M., et al. 2009, ApJ, 704, 496\\
Brown, J. M., et al. 2012, ApJ, 744, 116\\
Bruderer, S., et al. 2014, A\&A, 562, A26\\
Carrasco-Gonzalez, C., et al. 2016, ApJ, 821, L16\\
Casassus, S., et al. 2013, Nature, 493, 191\\
Casassus, S., et al. 2015, ApJ, 812, 126\\
de Val-Borro, M., et al. 2007, A\&A, 471, 1043\\
Dong, R., et al. 2017, ApJ, 836, 201\\
Dong, R., et al. 2018, ApJ, 860, 124\\
Draine, B. T. 2006, ApJ, 636, 1114\\
Espaillat, C., et al. 2014, Protostars and Planets VI, 497\\
Fedele, D., et al. 2017, A\&A, 600, A72\\
Fuente, A., et al. 2017, ApJ, 846, L3\\
Garufi, A., et al. 2013, A\&A, 560, A105\\
Geers, V. C., et al. 2007, A\&A, 469, L35\\
Helled, R., et al. 2014, Protostars and Planets VI, 643\\
Isella, A., Carpenter, J. M., \& Sargent, A. I. 2009, ApJ, 701, 260 \\
Isella, A., et al. 2010, ApJ, 725, 1735\\
Isella, A., et al. 2016, Physical Review Letters, 117, 251101 \\
Johansen, A., Youdin, A., \& Klahr, H. 2009, ApJ, 697, 1269 \\
Kessler-Silacci, J., et al. 2006, ApJ, 639, 275\\
Klahr, H. H., \& Henning, T. 1997, Icarus, 128, 213\\
Kraus, S., et al. 2017, ApJ, 848, L11\\
Lin, D. N. C., \& Papaloizou, J. 1979, MNRAS, 188, 191 \\
Lovelace, R. V. E., et al. 1999, ApJ, 513, 805\\
Lyra, W., et al. 2009, A\&A, 493, 1125\\
Marino, S., Perez, S., \& Casassus, S. 2015, ApJ, 798, L44\\
Natta, A., et al. 2007, Protostars and Planets V, 767\\
Okuzumi, S., et al. 2016, ApJ, 821, 82\\
Panic, O., \& Hogerheijde, M. R. 2009, A\&A, 508, 707\\
Perez, L. M., et al. 2014, ApJ, 783, L13\\
Perez, L. M., et al. 2015, ApJ, 813, 41\\
Perez, S., et al. 2015, ApJ, 798, 85\\
Pinilla, P., Benisty, M., \& Birnstiel, T. 2012, A\&A, 545, A81 \\
Pinilla, P., et al. 2015, A\&A, 573, A9\\
Pinilla, P., et al. 2017a, ApJ, 839, 99\\
Pinilla, P., et al. 2017b, ApJ, 839, 99\\
Ragusa, E., et al. 2017, MNRAS, 464, 1449\\
Reg?aly, Z., et al. 2012, MNRAS, 419, 1701\\
Ricci, L., et al. 2010, A\&A, 512, A15\\
Rice, W. K. M., et al. 2006, MNRAS, 373, 1619\\
Takahashi, S. Z., \& Inutsuka, S.-i. 2016, AJ, 152, 184\\
Tazzari, M., et al. 2016, A\&A, 588, A53\\
Testi, L., et al. 2003, A\&A, 403, 323\\
Testi, L., et al. 2014, Protostars and Planets VI, 339\\
Tripathi, A., et al. 2017, ApJ, 845, 44\\
Tsukagoshi, T., et al. 2016, ApJ, 829, L35\\
van Boekel, R., et al. 2003, A\&A, 400, L21\\
van der Marel, N., et al. 2013, Science, 340, 1199\\
van der Marel, N., et al. 2015a, ApJ, 810, L7\\
van der Marel, N., et al. 2015b, A\&A, 579, A106\\
van der Marel, N., et al. 2016a, ApJ, 832, 178\\
van der Marel, N., et al. 2016b, A\&A, 585, A58\\
van der Marel, N., et al. 2018, ApJ, 854, 177\\
van der Plas, G., et al. 2017, A\&A, 607, A55\\
Varniere, P., \& Tagger, M. 2006, A\&A, 446, L13\\
Weidenschilling, S. J. 1977, MNRAS, 180, 57\\
Whipple, F. L. 1972, in From Plasma to Planet, 211 \\
Williams, J. P., \& Cieza, L. A. 2011, ARA\&A, 49, 67\\
Winn, J. N., \& Fabrycky, D. C. 2015, ARA\&A, 53, 409\\
Wolf, S., \& Klahr, H. 2002, ApJ, 578, L79\\
Zhang, K., Blake, G. A., \& Bergin, E. A. 2015, ApJ, 806, L7 \\
Zhang, K., et al. 2014, ApJ, 791, 42\\
Zhu, Z., et al. 2012, ApJ, 755, 6\\


\normalsize

\end{document}